\documentclass[twocolumn,showpacs,preprintnumbers,amsmath,amssymb,prl]{revtex4}

\usepackage{graphicx}
\usepackage{dcolumn}
\usepackage{bm}

\begin{document}

\title{High-fidelity teleportation beyond the no-cloning limit and entanglement swapping\\
for continuous variables}

\author{Nobuyuki Takei$^{1,2}$, Hidehiro Yonezawa$^{1,2}$, Takao Aoki$^{1,2}$, and Akira Furusawa$^{1,2}$}

\affiliation{%
$^{1}$Department of Applied Physics, School of Engineering, The University of Tokyo,\\
7-3-1 Hongo, Bunkyo-ku, Tokyo 113-8656, Japan\\
$^{2}$CREST, Japan Science and Technology (JST) Agency, 1-9-9 Yaesu, Chuo-ku, Tokyo 103-0028, Japan
}%

\date{\today}

\begin{abstract}

We experimentally demonstrate continuous-variable quantum teleportation beyond the no-cloning limit. 
We teleport a coherent state and achieve the fidelity of 0.70$\pm$0.02 that surpasses the no-cloning limit of 2/3.  
Surpassing the limit is necessary to transfer the nonclassicality of an input quantum state. 
By using our high-fidelity teleporter, we demonstrate entanglement swapping, namely teleportation of quantum entanglement, as an example of transfer of nonclassicality. 

\end{abstract}


\maketitle

Quantum teleportation \cite{Bennett93,Bouwmeester97,Braunstein98,vanLoock00} is an essential protocol in quantum communication and quantum information processing \cite{Braunstein03,Nielsen00}. 
This protocol enables reliable transfer of an arbitrary, unknown quantum state from one location to another. 
This transfer is achieved by utilizing shared quantum entanglement and classical communication between two locations. 
Experiments of quantum teleportation have been successfully demonstrated with photonic qubits \cite{Bouwmeester97} and atomic qubits \cite{Riebe04,Barrett04} and also realized in optical field modes \cite{Furusawa98,Bowen03,Zhang03}. 
In particular, the teleportation experiments with atomic qubits and optical field modes are considered to be deterministic or unconditional.

Quantum teleportation can also be combined with other operations to construct advanced quantum circuits in quantum information processing \cite{Braunstein03,Nielsen00}. 
The teleported state will be manipulated in subsequent operations, some of which may rely on the nonclassicality contained in the state.
Therefore it is desirable to realize a high-quality teleporter which preserves the nonclassicality throughout the process.

In a continuous-variable (CV) system \cite{Braunstein98,vanLoock00}, a required quality to accomplish the transfer of nonclassicality is as follows: the fidelity $F_c$ of a coherent state input exceeds 2/3 at unity gains of classical channels \cite{Ban04}. 
Here the fidelity is a measure that quantifies the overlap between the input and the output states: $F=\langle \psi_{in} |\hat{\rho}_{out} |\psi_{in} \rangle$ \cite{Braunstein00}. 
Quantum teleportation succeeds when the fidelity exceeds the classical limit ($F_c=1/2$ for a coherent state input) which is the best achievable value without the use of entanglement. 
The value of 2/3 is referred to as the no-cloning limit, because surpassing this limit warrants that the teleported state is the best remaining copy of the input state \cite{Grosshans01}. 
As mentioned at the beginning, the essence of teleportation is the transfer of an arbitrary quantum state. 
To achieve it, the gains of classical channels must be set to unity. 
Otherwise the displacement of the teleported state does not match that of the input state, and the fidelity drops to zero when it is averaged over the whole phase space \cite{vanLoock00}. 
Note that the concept of gain is peculiar to a CV system and there is no counterpart in a qubit system.

A teleporter surpassing the no-cloning limit enables the transfer of the following nonclassicality in an input quantum state. 
It is possible to transfer a negative part of the Wigner function of a quantum state like the Schr\"odinger-cat state $|\psi_{cat} \rangle \propto |\alpha \rangle \pm |-\alpha \rangle$ and a single photon state \cite{Ban04}. 
The negative part is the signature of the nonclassicality \cite{Leonhardt97}. 
Moreover, two resources of quantum entanglement for teleporters surpassing the no-cloning limit allows one to perform entanglement swapping \cite{Pan98,Tan99}: one resource of entanglement can be teleported by the use of the other. 
The teleported entanglement is still capable of bipartite quantum protocols (e.g., quantum teleportation).

Although quantum teleportation of coherent states has been successfully performed \cite{Furusawa98,Zhang03,Bowen03} and the fidelity $F_c$ beyond the classical limit of 1/2 \cite{Braunstein00} has been obtained, $F_c > 2/3$ has never been achieved.
In terms of the transfer of nonclassicality, entanglement swapping has been demonstrated recently \cite{Jia04}. 
However, the gains of classical channels were tuned to optimal values (non-unity) for the transfer of the particular entanglement. 
At such non-unity gains, one would fail in teleportation of other input states such as a coherent state.

In this Letter we demonstrate teleportation of a coherent state at unity gains, and we achieve the fidelity of $0.70\pm0.02$ surpassing $F_c =2/3$ for the first time to the best of our knowledge. 
By using our teleporter we demonstrate entanglement swapping as an example of teleportation of nonclassicality. 
The gains of our teleporter are always set to unity to teleport an arbitrary state. 

The quantum state to be teleported in our experiment is that of an electromagnetic field mode as in the previous works \cite{Furusawa98,Zhang03,Bowen03,Jia04}. 
An electromagnetic field mode is represented by an annihilation operator $\hat{a}$ whose real and imaginary parts ($\hat{a}=\hat{x}+i\hat{p}$) correspond to quadrature-phase amplitude operators with the canonical commutation relation $[\hat{x}, \hat{p}]=i/2$ (units-free, with $\hbar=1/2$).

The fidelity $F_c$ is mainly limited by the degree of correlation of shared quantum entanglement between sender Alice and receiver Bob. 
For CVs such as quadrature-phase amplitudes, the ideal EPR (Einstein-Podolsky-Rosen) entangled state shows entanglement of $\hat{x}_{i}-\hat{x}_{j} \to 0$ and $\hat{p}_{i}+\hat{p}_{j} \to 0$, where subscripts $i$ and $j$ denote two relevant modes of the state. 
The existence of entanglement between the relevant modes can be checked by the inseparability criterion \cite{Duan00,Simon00}: $\Delta_{i,j} \equiv \langle [\Delta (\hat{x}_{i}-\hat{x}_{j} )]^2\rangle + \langle [\Delta (\hat{p}_{i} +\hat{p}_{j})]^2\rangle <1$, where the variances of a vacuum state are $\langle (\Delta \hat{x}^{(0)})^2 \rangle=\langle (\Delta \hat{p}^{(0)})^2 \rangle=1/4$ and a superscript (0) denotes the vacuum state. 
If this inequality holds, the relevant modes are entangled. 
In the case in which Alice (mode A) and Bob (mode B) share entanglement of $\langle [\Delta (\hat{x}_{\mathrm{A}}-\hat{x}_{\mathrm{B}} )]^2\rangle \simeq \langle [\Delta (\hat{p}_{\mathrm{A}} +\hat{p}_{\mathrm{B}})]^2\rangle$, the inseparability criterion $\Delta_{\mathrm{A},\mathrm{B}}<1$ corresponds to the fidelity $F_c >1/2$ for a teleporter without losses \cite{Braunstein01}. 
Furthermore $\Delta_{\mathrm{A},\mathrm{B}}<1/2$ corresponds to the fidelity $F_c >2/3$. 
Therefore, in order to achieve $F_c >2/3$, we need quantum entanglement with at least $\Delta_{\mathrm{A},\mathrm{B}}<1/2$.

When $F_c >2/3$ is achieved, it is possible to perform entanglement swapping with the teleporter and an entanglement resource with $\Delta_{ref,in}<1/2$, where we assume that the entangled state consists of two sub-systems: `reference' and `input'. 
While the reference is kept during a teleportation process, the input is teleported to an output station. 
After the process, the success of this protocol is verified by examining quantum entanglement between the reference and the output: $\Delta_{ref,out}<1$. 
Note that to accomplish this protocol, we need two pairs of entangled states with $\Delta_{i,j}<1/2$.

\begin{figure}[t]
\scalebox{1.0}{\includegraphics{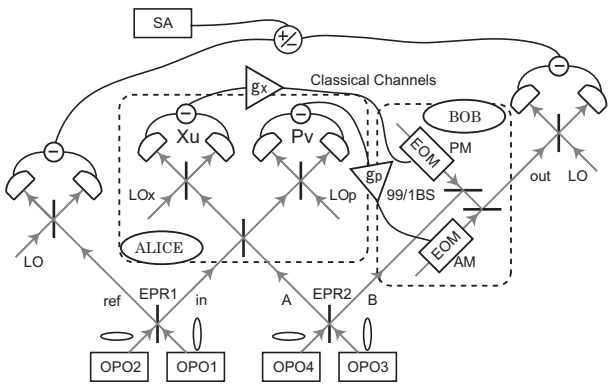}}
\caption{
The experimental set-up for teleportation of quantum entanglement. 
OPOs are optical parametric oscillators. 
All beam splitters except 99/1 BSs are 50/50 beam splitters. 
LOs are local oscillators for homodyne detection. 
SA is a spectrum analyzer. 
The ellipses illustrate the squeezed quadrature of each beam. 
Symbols and abbreviations are defined in the text.
}
\end{figure}
The scheme for entanglement swapping is illustrated in Fig. 1. 
Two pairs of entangled beams denoted by EPR1 and EPR2 are generated by combining squeezed vacuum states at half beam splitters. 
One of the EPR1 beams is used as a reference. 
The other is used as an input and teleported to the output mode. 
The EPR2 beams consist of mode A and B, and they are utilized as a resource of teleportation. 
In the case of a coherent state input, a modulated beam is put into the input mode instead of the EPR1 beam. 

Each squeezed vacuum state is generated from a subthreshold optical parametric oscillator (OPO) with a potassium niobate crystal (length 10mm). 
The crystal is temperature-tuned for type-I noncritical phase matching. 
Each OPO cavity is a bow-tie-type ring cavity which consists of two spherical mirrors (radius of curvature 50 mm) and two flat mirrors. 
The round trip length is 500 mm and the waist size in the crystal is 20$\mu$m. 
The output of a continuous wave Ti:Sapphire laser at 860nm is frequency doubled in an external cavity with the same configuration as the OPOs. 
The output beam at 430nm is divided into four beams to pump four OPOs. 
The pump power is about 80mW for each OPO.

We describe here a teleportation process in the Heisenberg picture. 
First Alice and Bob share entangled EPR2 beams of mode A and B. 
Alice performs ``Bell measurement" on her entangled mode ($\hat{x}_{{\rm A}} ,\hat{p}_{{\rm A}}$) and an unknown input mode ($\hat{x}_{in} ,\hat{p}_{in}$). 
She combines these modes at a half beam splitter and measures $\hat{x}_u=(\hat{x}_{in}-\hat{x}_{\rm{A}})/\sqrt{2}$ and $\hat{p}_v=(\hat{p}_{in}+ \hat{p}_{\rm{A}})/\sqrt{2}$ with two optical homodyne detectors. 
These measured values $x_u$ and $p_v$ for $\hat{x}_u$ and $\hat{p}_v$ are sent to Bob through classical channels with gains $g_x$ and $g_p$, respectively.

The gains are adjusted in the manner of Ref.~\cite{Zhang03}. 
The normalized gains are defined as 
$g_x =\langle \hat{x}_{out} \rangle /\langle \hat{x}_{in} \rangle$ and $g_p =\langle \hat{p}_{out} \rangle /\langle \hat{p}_{in} \rangle$. 
We obtain the measured gains of $g_x =1.00 \pm0.02$ and $g_p =0.99 \pm0.02$, respectively. 
For simplicity, these gains are fixed throughout the experiment and treated as unity.

Let us write Bob's initial mode before the measurement of Alice as: $\hat{x}_{{\rm B}} = \hat{x}_{in} -( \hat{x}_{\rm A}-\hat{x}_{\rm B})-\sqrt{2} \hat{x}_{u}$ and $\hat{p}_{{\rm B}} = \hat{p}_{in} +(\hat{p}_{\rm A}+\hat{p}_{\rm B})-\sqrt{2} \hat{p}_{v}$. 
Note that in this step Bob's mode remains unchanged. 
After measuring $\hat{x}_{u}$ and $\hat{p}_{v}$ at Alice, these operators collapse and reduce to certain values. 
Receiving her measurement results, Bob displaces his mode as 
$\hat{x}_{{\rm B} } \to \hat{x}_{out}=\hat{x}_{{\rm B} } + \sqrt{2} g_x x_{u}$,\ $\hat{p}_{{\rm B} } \to \hat{p}_{out}=\hat{p}_{{\rm B} } + \sqrt{2} g_p p_{v}$ and accomplishes the teleportation. 
Here we write explicitly the gains $g_x$ and $g_p$ to show the meaning of them, but they are treated as unity as mentioned before. 
In our experiment, displacement operation is performed by using electro-optical modulators (EOMs) and highly reflecting mirrors (99/1 beam splitters). 
Bob modulates two beams by using amplitude and phase modulators (AM and PM in Fig. 1). 
We use two beams to avoid the mixing of amplitude and phase modulations. 
The amplitude and phase modulations correspond to the displacement of $p$ and $x$ quadratures, respectively. The modulated beams are combined with Bob's mode ($\hat{x}_{\rm B} ,\hat{p}_{\rm B}$) at 99/1 beam splitters.

The teleported mode becomes
\begin{eqnarray}
\hat{x}_{out} &=& \hat{x}_{in} - (\hat{x}_{\rm A}-\hat{x}_{\rm B}), \nonumber  \\
\hat{p}_{out} &=& \hat{p}_{in} + (\hat{p}_{\rm A}+\hat{p}_{\rm B}).
\label{eq:output}
\end{eqnarray}
In the ideal case, the EPR2 state is the state for which $\hat{x}_{\rm A}-\hat{x}_{\rm B} \to 0$ and $\hat{p}_{\rm A}+\hat{p}_{\rm B} \to 0$. 
Then the teleported state is identical to the input state.
In real experiments, however, the teleported state has additional fluctuations. 
Without entanglement, at least two units of vacuum noise are added \cite{Braunstein98}. 
In other words, the noise $\langle [\Delta (\hat{x}_{\mathrm{A}}-\hat{x}_{\mathrm{B}} )]^2\rangle \ge 2\times\frac{1}{4}$ is added in $x$ quadrature (similarly in $p$ quadrature). 
These variances correspond to $\Delta_{\mathrm{A},\mathrm{B}}\ge 1$, resulting in the fidelity $F_c \le 1/2$. 
On the other hand, with entanglement, added noise is less than two units of vacuum noise. 
In the case with entanglement of $\Delta_{\mathrm{A},\mathrm{B}}<1/2$ which is necessary to accomplish $F_c >2/3$, the added noise is less than a unit of vacuum noise.

\begin{figure}[t]
\scalebox{0.4}{\includegraphics{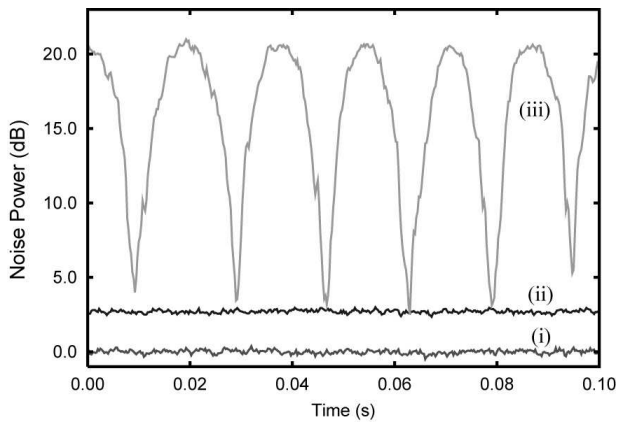}}
\caption{
The measurement results of the teleported state for a coherent state input in $x$ quadrature. 
Each trace is normalized to the corresponding vacuum noise level. 
Trace i shows the corresponding vacuum noise level $\langle (\Delta \hat{x}_{out}^{(0)} )^2\rangle =1/4$. 
Traces ii shows the teleported state for a vacuum input. 
Note that the variance of the teleported state for a vacuum input corresponds to that for a coherent state input. 
Trace iii shows the teleported state for a coherent state input with the phase scanned. 
At the top (bottom) of the trace, the relative phase between the input and the LO is $0$ or $\pi$ ($\pi/2$ or 3$\pi$/2). 
The measurement frequency is centered at 1 MHz, and the resolution and video bandwidths are 30kHz and 300 Hz, respectively. 
Traces i and ii are averaged 20 times.
}
\end{figure}
We first perform teleportation of a coherent state to quantify the quality of our teleporter with the fidelity $F_c$. 
In our experiment, we use frequency sidebands at $\pm$1MHz of an optical carrier beam as a quantum state. 
Thus a coherent state can be generated by applying phase modulation with EOM to the carrier beam. 
This modulated beam is put into the input mode instead of the EPR1 beam. 
Figure 2 shows measurement results of the teleported mode. 
The measured amplitude of the coherent state is $20.7\pm 0.2 $dB compared to the corresponding vacuum noise level. 
The measured values of the variances are $\langle (\Delta \hat{x}_{out} )^2 \rangle =2.82 \pm 0.09$dB and $\langle (\Delta \hat{p}_{out} )^2 \rangle =2.64\pm 0.08$ dB (not shown). 
The fidelity for a coherent state input can be written as $F_c =2/\sqrt{(1+4\sigma_x)(1+4\sigma_p)}$, where $\sigma_x =\left \langle (\Delta \hat{x}_{out} )^2\right \rangle$ and $\sigma_p =\left \langle (\Delta \hat{p}_{out} )^2\right \rangle$ \cite{Furusawa98,Braunstein01}. 
The fidelity obtained from the measured variances is $F_c =0.70 \pm 0.02$. 
This result clearly shows the success of teleportation of a coherent state beyond the no-cloning limit. 
Moreover we examine the correlation of the EPR2 beams and obtain the entanglement of $\Delta_{{\rm A},{\rm B}}=0.42 \pm 0.01$, from which the expected fidelity of $F_c =0.70\pm0.01$ is calculated. 
The experimental result is in good agreement with the calculation. 
Such good agreement indicates that our phase-locking system is very stable and that the fidelity is mainly limited by the degree of entanglement of the resource. 
As discussed in Ref.~\cite{Zhang03}, residual phase fluctuation in a locking system affects an achievable fidelity, and probably has prevented previous works from surpassing the no-cloning limit. 
Highly stabilized phase-locking system (both mechanically and electronically) allows us to achieve the fidelity of 0.70.

\begin{figure}[t]
\scalebox{0.4}{\includegraphics{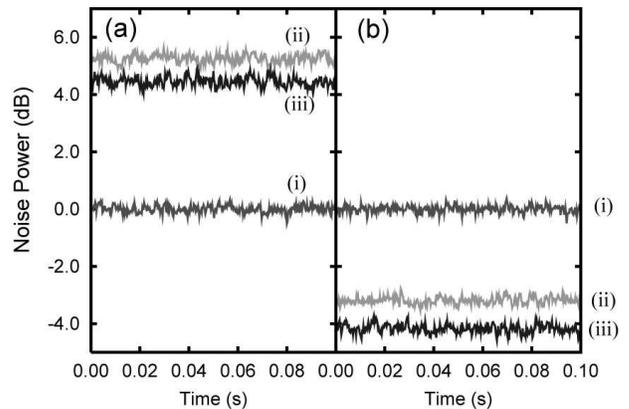}}
\caption{
Correlation measurement for EPR1 beams. 
(a) The measurement result of the reference mode alone. 
Trace i shows the corresponding vacuum noise level $\langle (\Delta \hat{x}_{ref}^{(0)} )^2\rangle =\langle (\Delta \hat{p}_{ref}^{(0)} )^2\rangle =1/4$. 
Traces ii and iii are the measurement results of $\langle (\Delta \hat{x}_{ref} )^2\rangle$ and $\langle (\Delta \hat{p}_{ref} )^2\rangle$, respectively. 
(b) The measurement result of the correlation between the input mode and the reference mode. 
Trace i shows the corresponding vacuum noise level $\langle [\Delta (\hat{x}_{ref}^{(0)}-\hat{x}_{in}^{(0)} )]^2\rangle =\langle [\Delta (\hat{p}_{ref}^{(0)} +\hat{p}_{in}^{(0)})]^2\rangle =1/2$. 
Traces ii and iii are the measurement results of $\langle [\Delta (\hat{x}_{ref}-\hat{x}_{in} )]^2\rangle$ and $\langle [\Delta (\hat{p}_{ref} +\hat{p}_{in})]^2\rangle$, respectively. 
The measurement condition is the same as that of Fig. 2.
}
\end{figure}
Next we demonstrate entanglement swapping. 
Before performing the experiment, we measure the noise power of each mode for EPR1 beams and the initial correlation between the modes with homodyne detection. 
For the reference mode, we obtain the noise levels of $5.23 \pm 0.14$dB and $4.44 \pm 0.14$dB for $x$ and $p$ quadratures, respectively (Fig. 3a). 
Similarly, the noise levels of $5.19 \pm 0.13$dB and $4.37 \pm 0.14$dB are obtained for $x$ and $p$ quadratures for the input mode (not shown). 
By making electrical subtraction or summation of the homodyne detection outputs, we observe the noise levels of $-3.19 \pm 0.13$dB for $x$ quadrature and $-4.19 \pm 0.14$dB for $p$ quadrature (Fig. 3b). 
From these values, we obtain the measured variances of $\Delta_{ref,in}=0.43 \pm 0.01<1$. 
This result shows the existence of the quantum entanglement between the input and the reference, and also indicates that we can transfer this entanglement with our teleporter. 

\begin{figure}[t]
\scalebox{0.4}{\includegraphics{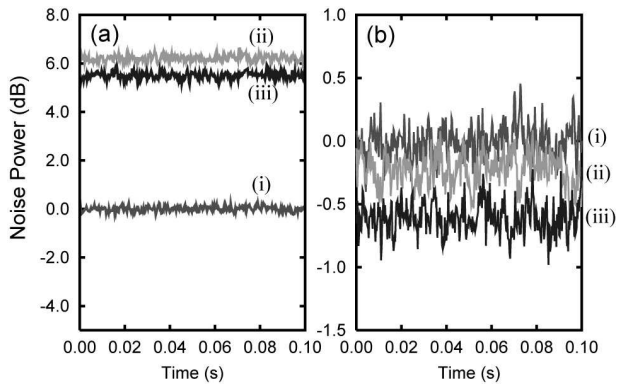}}
\caption
{
Correlation measurement results of the teleportation of quantum entanglement. 
(a) The measurement result of the output mode alone. 
Trace i shows the corresponding vacuum noise level $\langle (\Delta \hat{x}_{out}^{(0)} )^2\rangle =\langle (\Delta \hat{p}_{out}^{(0)} )^2\rangle =1/4$. 
Traces ii and iii are the measurement results of $\langle (\Delta \hat{x}_{out} )^2\rangle$ and $\langle (\Delta \hat{p}_{out} )^2\rangle$, respectively. 
(b) The measurement result of the correlation between the output mode and the reference mode. 
Trace i shows the corresponding vacuum noise level $\langle [\Delta (\hat{x}_{ref}^{(0)}-\hat{x}_{out}^{(0)} )]^2\rangle =\langle [\Delta (\hat{p}_{ref}^{(0)} +\hat{p}_{out}^{(0)})]^2\rangle =1/2$. 
Traces ii and iii are the measurement results of $\langle [\Delta (\hat{x}_{ref}-\hat{x}_{out} )]^2\rangle$ and $\langle [\Delta (\hat{p}_{ref} +\hat{p}_{out})]^2\rangle$, respectively. 
The measurement condition is the same as that of Fig. 2.
}
\end{figure}
We then proceed to the experiment of entanglement swapping and measure the correlation between the output and the reference in a similar way. 
The state in the reference mode does not change in the process. 
For the output mode, the noise levels of $6.06 \pm 0.12$dB and $5.47 \pm 0.14$dB are obtained for $x$ and $p$ quadratures, respectively, as shown in Fig. 4a. 
Because of the imperfect teleportation, some noises are added to the teleported state, resulting in the larger variances than that of the reference. 
Figure 4b shows the results of the correlation measurement. 
We observe the noise levels of $-0.25 \pm 0.13$dB and $-0.60 \pm 0.13$dB for $x$ and $p$ quadratures, respectively, yielding $\Delta_{ref,out}=0.91 \pm 0.02<1$. 
This result clearly shows the existence of quantum entanglement between the output and the reference. 
Therefore we can declare the success of entanglement swapping with unity gains.

In summary, we have demonstrated teleportation of a coherent state with the fidelity of 0.70$\pm$0.02. 
By using this high-fidelity teleporter, we have demonstrated entanglement swapping, or teleportation of quantum entanglement as a nonclassical input. 
Moreover, this high-quality teleporter will allow us to apply the teleported state to the subsequent manipilations and the construction of an advanced quantum circuits. 
For example, a bipartite quantum protocol like quantum teleportation can be performed by using the swapped entanglement. 
In addition, our teleporter has the capability of transferring a negative part of the Wigner function of a quantum state like a single photon state.

This work was partly supported by the MEXT and the MPHPT of Japan, and Research Foundation for Opto-Science and Technology.

\end{document}